\documentstyle[12pt,arry]{article}
 \let\ssection=\section
\def\section{\setcounter{equation}{0}\ssection}

\font\openbig=msym10 scaled \magstep1
\font\openscr=msym8 scaled \magstep1
\font\openscrscr=msym8
\newfam\openfam
\textfont\openfam=\openbig
\scriptfont\openfam=\openscr
\scriptscriptfont\openfam=\openscrscr
\def\open{\fam\openfam}
\font\Scbig=cmss10 scaled\magstep1
\font\Scscr=cmss8 scaled\magstep1
\font\Scscrscr=cmss8
\newfam\Scfam
\textfont\Scfam=\Scbig
\scriptfont\Scfam=\Scscr
\scriptscriptfont\Scfam=\Scscrscr
\def\Sc{\fam\Scfam}

\def\req#1{(\ref{#1})}
\def\d{\partial}
\def\dd#1{{\partial\over\partial{#1}}}
\def\pr{^\prime}
\def\bar{\overline}

\def\BE{\begin{equation}}
\def\EE{\end{equation}}
\def\BA{\begin{array}}
\def\EA{\end{array}}
\def\L{\left}
\def\R{\right}

\def\half{{1\over2}}

\def\Kr#1{\delta_{{#1},0}}
\def\ket#1{|{#1}\rangle}
\def\ccases#1#2{\L\{\new\BA{l}{#1}\\ {#2}\EA\R.}

\def\hn{\hat{n}}
\def\hL{\widehat{L}}
\def\hT{\widehat{T}}
\def\cL{{\cal L}}
\def\cH{{\cal H}}
\def\cG{{\cal G}}
\def\cQ{{\cal Q}}
\def\oZ{{\open Z}}
\def\ctop{{\Sc c}}
\def\htop{{\Sc h}}

\def\a{\alpha}

\def\g{\gamma}

\def\de{decoupling equation}
\def\cs{constraints}
\def\K{Kontsevich}
\def\M{Miwa}
\def\V{Virasoro}
\def\h{hierarchy}
\def\hs{hierarchies}
\def\emt{energy-momentum tensor}
\def\cc{central charge}
\def\tcc{topological central charge}
\def\ie{{\rm i.e.\/}}
\def\lvm{\leavevmode\hbox to\parindent{\hfill}}
\begin{document}
\hfuzz=1pt
\title{{\sc $d\leq1\bigcup d\geq25$ and Constrained KP
 Hierarchy from BRST Invariance in
the c~$\neq3$ Topological Algebra}}\author{{\large {\bf B.~Gato-Rivera}}\\
{\small {\sl CERN, CH1211 Geneva 23, Switzerland}}\\
{\small and}\\ {\small
{\sl Instituto de Matem\'aticas
y F\'\i sica Fundamental, Serrano 123, Madrid
28006, Spain}}\\{}\\ and\\ {}\\ {\large {\bf A.~M.~Semikhatov}}\\ {\small
{\sl Theory Division, P.~N.~Lebedev Physics Institute}}\\ {\small {\sl
Leninsky prosp. 53, Moscow 117924, Russia}}} \date{\relax{}} \maketitle
\begin{abstract} The BRST invariance condition in a highest-weight
representation of the topological ($\equiv$ twisted $N\!=\!2$) algebra
captures the `invariant' content of two-dimensional gravity coupled to
matter. The topological algebra
allows reductions to either the DDK-dressed
matter or the `\K-\M'-dressed matter related to \V-constrained
KP \h. The standard DDK formulation
is recovered by splitting the topological
generators into $c=-26$ reparametrization ghosts + matter + `Liouville',
while a similar splitting involving $c=-2$ ghosts gives rise to the matter
dressed in exactly the way required
in order that the theory be equivalent to
\V\ \cs\ on the KP \h. The two dressings of matter with the `Liouville'
differ also by their `ghost numbers', which is similar to the existence of
representatives of BRST cohomologies
with different ghost numbers. The \tcc\
$\ctop\neq3$ provides a two-fold
covering of the allowed region $d\leq1\cup
d\geq25$ of the matter central charge $d$ via
$d\!=\!(\ctop+1)(\ctop+6)/(\ctop-3)$.
The `Liouville' field is identified as
the ghost-free part of the topological $U(1)$ current. The construction
thus allows one to establish a direct relation (presumably an equivalence)
between the \V-constrained KP hierarchies, minimal models, and the BRST
invariance condition for highest-weight states of the topological
algebra.\end{abstract}

\thispagestyle{empty}

\vfill{
\hbox to\hsize{
CERN-TH.6527/92\hfill June 1992}
\hbox to\hsize{
IMAFF-92/7\hfill}
\hbox to\hsize{
hepth@xxx/9207004\hfill}
\hbox to\hsize{{\sf Revised version}\hfill}}

\newpage

\setcounter{page}{1}

\section{Introduction}\lvm It has been one of the main results of the
development of matrix models
\cite{[BK],[DSh],[GM]} that appropriately
constrained \cite{[D],[FKN1],[DVV]}
integrable \hs\ provide a description of
two-dimensional gravity coupled to matter.
However, until recently a direct
relation between the `integrable' formulation and the continuum approach
\cite{[KPZ],[Da],[DK]}
(\ie\ a relation that would not require going through
the {\it matrix\/} formulation) was missing. By using the {\sl \K-\M\
transform\/} on the KP \h, one such relation can be established
\cite{[S35],[GS]}~\footnote{Of course, the idea of the \K-\M\ transform
\cite{[S35]} was motivated by the previous works on matrix models and, in
particular, on the \K\ model; see for instance refs.
\cite{[K]} -- \cite{[GN]}.}. Namely, there is an isomorphism
\BE\BA{r}\mbox{Virasoro-constrained}\\ \mbox{KP\
\h}\EA\longleftrightarrow\BA{l}\mbox{`KM'-dressed}\\
\mbox{\phantom{`}matter}\EA\label{isom}\EE under which the \V\
\cs\
become level-$2$ \de s in the $(p\pr,p)$ minimal model with $p\pr/p$
determined as shown below in
eq.~\req{pproverp}, provided the minimal model
operators are {\sl dressed\/} with an extra $U(1)$ `Liouville' current
according to a `\K-\M' recipe to be described below.

The purpose of this paper is two fold.
{\bf First}, we would like to clarify
the relation between the two alternative descriptions of quantum gravity
coupled to matter: \V-constrained integrable \hs\ and the continuum
conformal field-theoretic formalism. As we will see, the outcome of the
\K-\M\ transform of the constrained KP  hierarchy is not quite the DDK
formalism, but rather a `topological'
version of it. The difference amounts
to the way the matter is dressed with
the `Liouville' field, and also to the
value of the `Liouville' background charge.
{\bf Second}, a universal notion
unifying the DDK and `KM' dressings will be identified as the topological
symmetry. It will be our aim to point out that the `invariant' meaning of
 gravity + matter theories is captured by a topological (twisted $N=2$)
symmetry \cite{[W-top],[EY]}
with arbitrary \tcc\ $\ctop\!\neq\!3$, in the
sense that different splittings
of the topological symmetry generators into
`matter' and `ghost' parts give rise to the corresponding dressing
prescriptions. We will consider
the two reductions represented by the upper
arrows in the diagram \BE\new\BA{rlcrl} {}&{}&\mbox{{\footnotesize
$\BA{c}\ctop\neq3\ \mathrm{topological}\\ \mathrm{algebra}\EA $}}&{}&{}\\
{}&\swarrow&{}&\searrow&{}\\ \mbox{{\footnotesize $\BA{r}
\mbox{{\mathrm`KM'-dressed}}\\ \mathrm{matter}\EA $}}
 &{}&{}&{}&\mbox{{\footnotesize $\BA{l} \mbox{{\mathrm DDK-dressed}}\\
\mathrm{matter}\EA $}}\\ {}&\searrow&{}&\swarrow&{}\\
{}&{}&\mbox{{\footnotesize $d\leq1\cup d\geq25$
matter}}&{}&{}\EA\label{diagram}\EE
They amount to splitting away
$c=-2$ and $c=-26$ ghosts, and lead directly to
the `KM' dressing prescription
and the DDK formalism, respectively. The lower
arrows tell us that the matter part identified inside the matter +
`Liouville' theory, is eventually the same in both cases, being the
standard one given in refs. \cite{[BPZ]} -- \cite{[DF]}.

As we will see, the matter central
charge $d$ is expressed as a function of
the \tcc\ via
\BE d={(\ctop+1)(\ctop+6)\over\ctop-3}\label{d(c)}\EE
and therefore we have $d\!\leq\!1$
or $d\!\geq\!25$ as a consequence of the
particular `breakdown' of the topological symmetry. As $\ctop$ grows from
$-\infty$ to 3 and from 3 to
$+\infty$, each of the allowed values of $d$ is
taken twice, except for the extrema of the function \req{d(c)}
$(\ctop\!=\!-3,d\!=\!1)$ and $(\ctop\!=\!9,d\!=\!25)$. Thus the \tcc\
provides a `two-sheeted covering' of the allowed region of the matter
\cc\footnote{This would be interesting to consider in relation with the
(somewhat mysterious) equivalence
between the conventional gravity (coupled
to matter) and the topological gravity (coupled to topological matter)
\cite{[W-Dec89]} -- \cite{[VV-Apr90]}. In particular, restrictions on the
central charge do not seem to have an immediate analog in the topological
gravity.}.

By the \tcc, or the anomaly, we mean
the coefficient (the honest \cc\ of the
untwisted algebra) read off from the algebra \cite{[W-top],[EY]}
\BE\new\BA{lclclcl}
\L[\cL_m,\cL_n\R]&=&(m-n)\cL_{m+n}~,&\qquad&[\cH_m,\cH_n]&=
&{\ctop\over3}m\Kr{m+n}~,\\
\L[\cL_m,\cG_n\R]&=&(m-n)\cG_{m+n}~,&\qquad&[\cH_m,\cG_n]&=&\cG_{m+n}~,\\
\L[\cL_m,\cQ_n\R]&=&-n\cQ_{m+n}~,&\qquad&[\cH_m,\cQ_n]&=&-\cQ_{m+n}~,\\
\L[\cL_m,\cH_n\R]&=&\multicolumn{5}{l}{-n\cH_{m+n}+{\ctop\over6}(m^2+m)
\Kr{m+n}~,}\\
\L\{\cG_m,\cQ_n\R\}&=&\multicolumn{5}{l}{2\cL_{m+n}-2n\cH_{m+n}+
{\ctop\over3}(m^2+m)\Kr{m+n}~,}\EA\label{topalgebra}\EE
while given by eq. \req{d(c)} is the usual matter \cc, \ie\ the permanent
hero of the square roots $\sqrt{(1-d)(25-d)}$. Note that the minimal-model
values of $d$,
\BE d=1-{6(p\pr-p)^2\over p\pr p}~,\label{dppr}\EE
come from the rational
values of the \tcc\footnote{So that at least formally
for $p\pr$ or $p=1$,
the corresponding formula ${\ctop\over3}=1-{2\over p}$
for the `topological' \cc\ becomes a particular case of the
``quasi-homogeneous"
formula from the Landau--Ginzburg description based on
the ideas from the catastrophe theory \cite{[M],[VW]}.}

\BE{\ctop\over3}=1-2\L({p\over p\pr}\R)^{\pm1}.\EE

To return to the diagrams \req{isom} and \req{diagram}, we recall
\cite{[S35]} that the `KM' prescription to dress matter with a
`Liouville', fixed by the machinery of the \K-\M\ transform, results in
level-$2$ null vector \de s  which
 turn out to be rather special with regard to the way
they involve (are dressed with)
the `Liouville' $U(1)$ current. Now, one may
wonder whether this special form of the \de s  admits an independent
explanation in the continuum theory.
As we will show, it is in fact governed
by a BRST invariance pertaining to a special realization of the algebra
\req{topalgebra}. This realization amounts to constructing the topological
symmetry generators in terms of
matter + `Liouville' + ghost theory with the
ghosts being of \cc\ $-2$.

We will show that the highest-weight conditions (including the BRST
invariance) w.r.t. the topological algebra reduce down to
matter~+~`Liouville' theory
to give our special form of the \de s, which, as
we know, characterizes the image of the \K-\M\ transform of the Virasoro
constraints. Thus, as a by-product,
the \K-\M\ transform is given the r\^ole
of a mapping from \V -constrained \hs\ to the highest-weight (and
BRST-invariance) conditions in the topological algebra. On the other hand,
the `true' Liouville field of the DDK formalism emerges as shown in the
diagram \req{diagram}: {\it the\/}
DDK formulation is recovered by a similar
reduction into matter + `Liouville' +
ghosts when the ghosts are taken to be
the usual $c\!=\!-26$ reparametrization ghosts.
Yet it is not this version,
but rather the one involving the $c\!=\!-2$ ghosts,
that we have been able to
relate to constrained integrable \hs. The `invariant' content of both
versions of matter dressed with an extra scalar
should thus be sought at the
level of the topological algebra, while the very fact that (at least) two
such versions make sense may be related to the
existence of the `non-standard
ghost-number' representatives of BRST cohomologies
\cite{[W-ground]} -- \cite{[LZ]}.

\section{\K-\M\ transform: from Virasoro \hfill\break
\cs\ to an `almost topological'
algebra}\lvm Classically, the \M\ transform is a presentation for the time
parameters $t_r$ of the KP hierarchy of the form \cite{[Mi],[Sa]}
\BE t_r={1\over r}\sum_j n_jz^{-r}_j,\quad r\geq1\label{Miwatransform}\EE
relating the (complexified) times to a set of points $\{z_j\}$ on the complex
plane\footnote{\label{infinitely}There have to be infinitely many points
$z_j$ in order that the times $t_r$ be independent.}. Very similar
parametrizations of the KP times have been used in recent works on the \K\
model cited above, hence the name. Applied to the Virasoro constrained
KP \h, the \K-\M\ transform allows one to recast the \cs\ imposed on the tau
function into the field-theoretic data, which are null vector \de s in a
conformal field theory on the $z$ plane \cite{[S35]}. The {\it\M\
parameters\/} $n_j$ are integer classically, but we will need to continue
them off the integer values, as they acquire the r\^ole of the `Liouville'
$U(1)$ charges.

Picking out a Miwa point $z_i$, one can {\sl solve\/} the Virasoro \cs
\BE\sum_{n\geq-1}\!\!z_i^{-n-2}{\Sc
L}_n\tau=0, \label{constraints}\EE
by substituting for the tau function the ansatz
\BE\tau(t)\equiv\tau\{z_j\}=\biggl\langle\Psi(z_i)\prod_{j\neq
i}\Psi(z_j)\biggr\rangle~.\label{ansatz}\EE
which allows one to establish
that the Virasoro generators acting on the tau
function get reformulated as a level-$2$ null vector decoupling operator
acting on the correlator.
The insertion at the point
$z_i$ is singled out in order to keep track of the
appearance of $z_i$ in the
\cs\ \req{constraints}. Clearly, these \cs\ can be
imposed for any of the $z_j$
and moreover, having done this for {\it every\/}
$z_j$, one gets an infinite (see footnote \ref{infinitely}) number of
equations on \req{ansatz}.

The conformal field theory understood on the RHS of \req{ansatz}
is constructed as follows \cite{[S35]}: One considers a semi-direct
product of the \V\ algebra with a $U(1)$ current, and thus introduces the
\emt\ $\hT(z)=\sum_{n\in {\open Z}}\hL_nz^{-n-2}$, and the $U(1)$ current
$I(z)=\sum_{n\in{\open Z}}I_n z^{-n-1}$.
Then, one restricts oneself to the
sector in which {\it all\/} the
(highest-weight) fields $\Psi_j$ have total
dimension zero and therefore are
only distinguished by their $U(1)$ charges
$n_j$ (which are the Miwa parameters!):
\BE\new\BA{rcl}\hL_0\ket{\Psi_j}&=&0,\qquad
I_0\ket{\Psi_j}~=~n_j\ket{\Psi_j},\\ \hL_{\geq1}\ket{\Psi_j}
&=&I_{\geq1}\ket{\Psi_j}~=~0\EA\label{highestweight}\EE
Further, the \cs\ imposed on the tau
 function are allowed to depend on a free
parameter, $J$, which can be viewed
 as the `spin' of an abstract
 $bc$ system\footnote{We would
 like to stress that, although the
 derivation of these constraints for arbitrary $J$ from a specific
{\it matrix\/} model may not be known, we postulate them and {\it then\/}
seek a relation with the continuum formalism, thereby promoting the
appropriately constrained \hs\ to a first principle of the `integrable'
description of quantum gravity
and bypassing the specifically {\it matrix\/}
formulation. Consistency of the Virasoro action via the generators
\req{Lontau} with the KP flows
has been proved long time ago in \cite{[S10]}
(see also \cite{[GO]} and references therein).}  :
\BE\new \BA{rcl} {\Sc L}_{p>0} &=&\half\sum^{p-1}_{s=1}{\d^{2}\over\d
t_{p-s}\d t_s}+\sum_{s\geq 1}st_s \dd{t_{p+s}}+\L(J-\half\R)(p+1)
\dd{t_p}\EA\label{Lontau}\EE
(for ${\Sc L}_0$ and ${\Sc L}_{-1}$ only the second term is present). The
value of $J$ fixes the $U(1)$ charge $n_i\equiv\hn$ of the insertion
$\Psi(z_i)$ at the selected point $z_i$, via
\BE2J-1={1\over \hn}-{2\hn}~.\label{*}\EE
$\Psi$ should be a $(2,1)$ (or $(1,2)$) operator in a $(p\pr, p)$ minimal
model \cite{[BPZ],[DF]},
{\it dressed\/} (so as to make its total dimension
zero) with the help of $\exp\L(\mbox{coeff.}\int^z\!I\R)$, for $p\pr/p$
determined by
\BE{p\pr\over
p}={2\hn^2}=1+{Q^2\over4}+{Q\over4}\sqrt{Q^2+8}~,\qquad
Q\equiv2J-1 .\label{pproverp}\EE
And finally, the generators introduced above satisfy,
\BE\new\BA{rcl}[I_m,I_n]&=&-m\Kr{m+n}\\
\L[\R.\!\hL_m,\hL_n]&=&(m-n)\hL_{m+n}+{2\over 12}(m^3 -m)\Kr{m+n}\\
\L[\R.\!\hL_m,I_n]&=&-nI_{m+n}- \half Q (m^2+m)\Kr{m+n}~.\EA\label{hat}\EE

The theory \req{hat},
emerging in this way from the \K-\M\ transform, will be
our starting point for constructing the topological generators. As a
motivation, notice that the algebra \req{hat} resembles the topological
algebra in the sense
that the matter \cc\ that one would expect according to
the presence of the
background charge $Q$, is `hidden' --- it does not show
up in the commutator between the \V\ generators. It will, however, appear
there if we go over to the reformulation in which the current is not
anomalous: introducing
\BE L_m=\hL_m- \half Q (m+1)I_m~,\label{shift}\EE
we find, instead of the corresponding formulae in \req{hat},
\BE\new\BA{rcl} \L[L_m,L_n\R]&=&(m-n)L_{m+n}+{d+1\over12}(m^3-m)\Kr{m+n}\\
\L[L_m,I_n\R]&=&-nI_{m+n}~,\EA\label{thetheory}\EE
with $d=1-3Q^2$. The $L_0$-dimensions $\Delta_j$ are no longer zero, but,
according to eq. \req{shift}, become proportional to the $U(1)$ charges
 $n_j$. This fixes the `KM' dressing prescription.

Further, subtracting away the $U(1)$ Sugawara contribution, by writing
\BE L_m=\bar{L}_m-\half\sum_{n\in\oZ}:I_{m-n}I_n:\label{barshift}\EE
we recover, in the $\bar{\phantom{Z}}$-sector, the standard minimal matter
 with \cc\ $d$ and, in particular, the `minimal' dimension of $\Psi$ :
\BE\new\BA{rcl}\delta&=&{3\over2}{\hn^2}-{1\over2}\\
{}&=&{5-d\pm\sqrt{(1-d)(25-d)}\over16}
\EA\label{delta}\EE
which is, of course,
the dimension of the corresponding $(2,1)$ (or $(1,2)$)
operator.

Thus the $U(1)$ current
might be considered superficial, as soon as it cannot
alter the `dynamical' content
of the story, which can only be based on {\it
the\/} \V\ null vectors. This is, however, precisely what we mean by {\sl
dressing\/}: the r\^ole of the $U(1)$ current is to rearrange the standard
\de s so as to make
them lie in the image of the \K-\M\ transform of the \V\
 \cs\ on the KP \h\footnote{Of course, the dressing by the $U(1)$
current contributes also to the conformal dimensions of the fields.}.

Namely,
after the \K-\M\ transform and
provided relation \req{*} holds,
the \V\ \cs\ $\sum_{n\geq-1}z_i^{-n-2}{\Sc L}_n\tau\!=\!0$
become \cite{[S35]}
\BE\L\{-{1\over 2n_i^2}
{\d^2\over\d z_i^2}+{1\over n_i}\sum_{j\neq i}{1\over
z_j-z_i}\L(n_j{\d\over\d z_i}-n_i{\d\over\d z_j}\R)\R\}
\biggl\langle\Psi(z_i)\prod_{j\neq
i}\Psi_j(z_j)\biggr\rangle=0~,\label{Tgen}\EE
where we have allowed correlation functions more general than
\req{ansatz}; recall that
the $n_j$ are the $U(1)$ charges of $\Psi_j$. This
equation can be interpreted as a \de\ corresponding to the null
vector\footnote{See also ref. \cite{[MS]} for a somewhat
different viewpoint of the relation between \V\ \cs\ and a particular case
$n_i^2\equiv n_j^2\equiv\half$
of (the `discrete analogue' of) eq. \req{Tgen}
for the Kontsevich partition function.}
\BE\new\BA{rcl}\ket\Upsilon&=&\L(\a\hL_{-1}^2+\hL_{-2}+\g
I_{-1}\hL_{-1}\R)\ket\Psi\\ \a&=&-{1\over2\hn^2}~,\qquad
\g~=~-{1\over\hn}~,\EA\label{UUpsilon}\EE
built upon the $\Psi$ operator.
A very special feature of this null vector is the absence of
the {\it a priori\/} possible ${I_{-1}}^2$ and $I_{-2}$ terms.
Looked at from the inside of the theory constructed solely out of
$\hL$ and $I$, this recipe to construct null vectors
(and hence the \de s) may seem somewhat {\it ad hoc\/}. As we
 are going to show, this is in fact inherited
from a BRST invariance, although there is no room for the latter in the
$\hT$-$I$ theory.

\section{BRST invariance: there and back again} \lvm As we have seen, the
\K-\M\ transform results in the minimal matter dressed with an extra
`Liouville' scalar in
a rather special way. Now, by introducing an auxiliary
$c=-2$ ghost system, we will further relate the thus dressed matter to a
topological theory in
which $I$ will become a part of the topological $U(1)$
current $\cH$ (see \req{topalgebra}). It will follow then that the special
form of the null vectors is governed by the BRST invariance
$\cQ(\ket\Upsilon\otimes\ket0)=0$ (where $\ket0$ is a ghost vacuum).

In this section, we will first show how the theory we dealt with in the
previous section can be embedded into a realization of the algebra
\req{topalgebra}. Then, from the BRST-invariance condition in a
highest-weight
representation of the topological algebra, we will reproduce
the `dressed' null vectors such as eq. \req{UUpsilon}.

Comparing \req{hat}
with the corresponding commutators from \req{topalgebra},
we observe an excess
of $+2$ of the \cc. Thus the good old $c=-2$ $bc$ system
\cite{[Di]} comes into play. Introducing the spin-1 (anticommuting) $bc$
system,
\BE\new\BA{l}b(z)=\sum_{n\in{\open Z}}b_nz^{-n-1},\qquad
c(z)=\sum_{n\in{\open Z}}c_nz^{-n},\\ \{b_n,c_m\}=\Kr{m+n}~,\qquad
b_{>-1}\ket{0}=c_{\geq1}\ket0=0~,\EA\label{spin1}\EE
will allow us to construct the topological algebra \req{topalgebra}.

The centerless \V\ generators entering \req{topalgebra} can be constructed
as, \BE\cL_m=\hL_m+l_m,\quad l_m\equiv\sum_{n\in{\oZ}}n:b_{m-n}c_n:
\label{L}\EE (the ghost \emt\ being $t(z)=-\!:\!b\d c\!:\!(z)$).  Further,
introducing the ghost
current $i=-:bc:$, we define the topological current as
\BE\cH_m=i_m+\sqrt{{3-\ctop\over3}}I_m~.\label{H}\EE However, as the ghost
field $b$ is of dimension 1, the more standard construction of the BRST
current $\cQ\sim cT$,
which works for $c$ of dimension $-1$, does not apply
here. Instead, we can
identify a dimension-1 odd current $\cQ(z)$ simply as
$\cQ(z)=b(z)$:  \BE\cQ_m=b_m\label{Q}~.\EE On the other hand, it is the
dimension-2
fermionic field $\cG(z)$ that now comprises all the `non-trivial'
terms usually characteristic of BRST generators built using a spin-2 $b$
field:

\BE\cG_m=2\sum_{p\in\oZ}c_{m-p}\hL_p+2\sqrt{{3-\ctop\over3}}
\sum_{p\in\oZ}(m-p)c_{m-p}I_p
+\!\sum_{p,r\in\oZ}(p+r-m):\!b_{p-r}c_rc_{m-p}\!:
+{\ctop\over3}(m^2+m)c_m\label{G}\EE

The coefficient in front of the second term, as it stands, is real for
$\ctop<3$; clearly, going over to $\ctop>3$ amounts to reversing the
signature of the fields.

The commutation relations \req{topalgebra} can now be checked easily, {\it
resulting, in particular, in the relation\/} \req{d(c)} {\it between the
matter and the \tcc s.\/}

We have thus arrived at a realization of the topological algebra
\req{topalgebra}. Now we will use this realization in order to establish
 that the image of
 the \K-\M\ transform of the constrained KP \h\ (\ie\ the
appropriately dressed \de s) can alternatively be described via the
BRST-invariance in a highest-weight representation of the topological
algebra. In particular,
we will show that the BRST-invariance condition in a
representation of the algebra \req{topalgebra}, gives rise, under the
splitting of a BRST-invariant highest-weight state into
matter$\otimes$ghosts, to the matter null vector $\ket{\Upsilon}$ dressed
according to the `\K-\M' prescription.

To begin with, fix a BRST-invariant ``chiral primary state" $\ket\Phi$
(cf. \cite{[LVW]}) ,
\BE\new\BA{rclclcl} \cQ_0\ket\Phi&=&0,&{}&\cG_0\ket\Phi&=&0,\\
\cL_{\geq0}\ket\Phi&=&\cH_{\geq1}\ket\Phi&=
&\cG_{\geq1}\ket\Phi&=&\cQ_{\geq1}\ket\Phi~=~0,\\
\cH_0\ket\Phi&=&\htop\ket\Phi.&{}&{}&{}&{}\EA\EE
The topological $U(1)$ charge $\htop$ is thus the only non-zero parameter
that distinguishes between the states. The BRST-invariant vectors,
$\cQ_0\ket\Xi=0$, at level 2, are of the form
\BE\ket{\Xi}=\L(\a\cL_{-1}^2+\cL_{-2}+\Gamma\cH_{-1}\cL_{-1}+
\half\Gamma\cQ_{-1}\cG_{-1}\R)\ket{\Phi}~,\label{Xi}\EE
with $\a$ and
$\Gamma$ arbitrary so far. An important point is that it is the
BRST-invariance
that rules out the {\it a priori\/} possible $\cH_{-2}$ and
$\cH_{-1}^2$ terms.

Imposing further the highest-weight conditions
\BE\cQ_{\geq1}\ket\Xi=\cG_{\geq1}\ket\Xi
=\cL_{\geq1}\ket\Xi=\cH_{\geq1}\ket\Xi=0~,\label{hwconditions}\EE
we arrive at the quadratic equation
\BE
2\htop^2-\htop\L({\ctop\over3}+1\R)+{\ctop\over3}-1=0~,
\label{hequation}\EE
and the corresponding solutions
\BE\htop=\ccases{\ctop-3\over6}{1},\qquad\a=
\ccases{6\over\ctop-3}{\ctop-3\over6},
\qquad\Gamma=\ccases{6\over3-\ctop}{-1}.\label{coefficients}\EE

Now, a crucial observation is that the state $\ket\Xi$ thus constructed is
not only BRST invariant, but also BRST-{\it exact\/}.
Such states should be
{\it identified\/} with zero (\ie\ factored out), which in our case would
mean imposing certain equations on the correlation functions involving
$\Phi$~\footnote{Clearly, it is not until the BRST
invariance is ensured in
this way for the correlators that one may expect
the latter to be independent
of the locations of operator insertions.}. The procedure is directly
analogous to the
use of \de s to factor away null vectors from Verma modules
\cite{[BPZ],[DF]}.
Moreover, as we will see in a moment, the state $\ket\Xi$
becomes precisely the
null vector $\ket\Upsilon$ (which is dressed according
to the `\K-\M' prescription)
when the ghosts are split away according to eqs.
\req{L}, \req{H}, \req{G}. Therefore, {\it the origin of the dressing
prescription, as well as of the null vector as such, can be traced to the
topological algebra and the condition of BRST invariance.}

To see how the objects we met with
in the previous section are recovered, we
introduce the Miwa parameter $\hn\equiv n_i$ (to become equal to the
`Liouville' charge of $\Phi(z_i)$) by parametrizing the topological $U(1)$
charge $\htop$ as
\BE \htop=\sqrt{{3-\ctop\over3}}\hn~.\EE
Therefore, the Miwa parameter is given in terms of the \tcc\ by
\BE\hn=\cases{-\half\sqrt{{3-\ctop\over3}}\cr\sqrt{{3\over3-\ctop}}}
\label{cases}.\EE

Assuming also the
generators of \req{topalgebra} to be of the form given in
\req{L}, \req{H}, \req{Q}, and \req{G}, and writing
$\ket\Phi=\ket\Psi\otimes\ket0$, where $\ket0$ is the ghost vacuum, we see
that the ghost contributions cancel out, and therefore $\ket\Xi$ can be
expressed as
\BE\ket\Xi=\ket\Upsilon\otimes\ket0\EE
with
\BE\ket\Upsilon=\L(\a\hL_{-1}^2+\hL_{-2}+\sqrt{{3-\ctop\over3}}~\!\Gamma
I_{-1}\hL_{-1}\R)\ket\Psi,~\EE
which coincides
with \req{UUpsilon} in view of \req{coefficients}. Shifting
the \V\ generators
as in \req{shift} and, further, as in \req{barshift}, one
recovers the standard
null vector in the $(d,\delta)$ Verma module, with the
central charge $d$ given
by \req{d(c)}. As for the highest weight $\delta$,
one only needs to substitute the values of $\hn$ \req{cases}
in \req{delta} (note the correspondence between the two cases
in \req{coefficients} and \req{cases},
and the $\pm$ of the square root in \req{delta}).

The regions $\ctop<3$ and $\ctop>3$ are `mirrored' in terms of the reduced
theory, in that the matter and the `Liouville' ($I$) take the place of one
another:
all the coefficients in the above ansatz for the topological algebra
generators
can be kept real when going over from $\ctop<3$ to $\ctop>3$, by
reversing the signature of the fields.

\section{Recovering the DDK formalism\hfill\break from the topological
symmetry}\lvm A legitimate question to ask at this point is how the above
theory, which
we have derived by `breaking down' the topological symmetry, is
related to the formalism of refs. \cite{[Da],[DK]}. It is amusing that the
DDK formalism
is recovered when one considers instead of the ansatz \req{L},
\req{H}, \req{Q}, \req{G}, {\it another\/} reduction of the type of
\BE\mbox{topological\ algebra}\Longrightarrow\mbox{matter +
`Liouville'+ ghosts}~.\label{splitting}\EE
To see this,
let us split away from the topological generators a spin-2 ghost
system. To this
end, we write\footnote{We continue to use the same notations,
although they have actually changed once the ghost system is different.}
\BE \cL_m=\hL_m+l_m,\quad
l_m\equiv\sum_{n\in{\oZ}}(m+n):b_{m-n}c_n:\label{L26}\EE
Then the \cc\ read off from the $\hL\hL$ commutator is 26. The DDK {\it
dressing prescription\/} can be recovered as follows. Recall that for a
spin-$\lambda$ ghost system, the $SL_2$ invariant vacuum state $\ket0$ is
characterized by \cite{[FMS]}
\BE b_{>-\lambda}\ket0=c_{\geq\lambda}\ket0=0\EE
(a particular case of which we have already met in \req{spin1}). For the
reparametrization ghosts it is thus only the $c_{\geq2}$ out of the $c_n$
modes that annihilate the vacuum, which allows us to split the topological
algebra states as
\BE\ket\Phi=\ket\Psi\otimes c_1\ket0~.\EE
In view of \req{L26} and the above highest-weight conditions, this implies
that the $\hL$-dimension
of $\Psi$ is $\widehat\Delta=1$. Clearly, the same
holds for all the other
`topological'-highest-weight states $\Phi_j$ of {\it
zero $\cL$-dimension\/}:
splitting them as $\ket{\Phi_j}=\ket{\Psi_j}\otimes
c_1\ket0$, we arrive at the conditions
\BE\widehat\Delta_j=1~,\EE
which will fix the DDK dressing prescription as soon as the background
charges are known. Thus, to complete the derivation, let us first give the
expression for the BRST current modes:
\BE\cQ_m=2\sum_{p\in\oZ}c_{m-p}\hL_p
+\sum_{p,r\in\oZ}(p-r):\!b_{m-p-r}c_pc_r\!: -2\sqrt{{3-\ctop\over3}}m
\sum_{p\in\oZ}c_{m-p}I_p +{\ctop\over3}(m^2-m)c_m~.\label{Q2}\EE
Then, with the superpartner of the \emt\ being
\BE\cG_m=b_m~,\EE
we arrive at the topological current given by
\BE\cH_m=\sum_{n\in\oZ}:\!b_{m-n}c_n\!:+\sqrt{{3-\ctop\over3}}I_m~.\EE
{}From the $[\cL_m,\cH_n]$ commutator we can now derive
\BE [\hL_m,I_n]=-nI_{m+n}+\sqrt{{3\over3-\ctop}}{\ctop-9\over6}(m^2+m)
\Kr{m+n}~.\label{hat26}\EE
By using \req{d(c)} the anomaly thus emerging can be expressed as
\BE\sqrt{{3\over3-\ctop}}{\ctop-9\over6}=-\half\sqrt{{25-d\over3}}
\equiv-\half Q_{{\mathrm L}}\EE
\ie\ it
coincides with the background charge $Q_{{\mathrm L}}$ of the standard
Liouville scalar.

\section{Conclusions:\hfill\break Virasoro
 \cs\ as BRST-invariance conditions}\lvm
We have
addressed the problem of how the constrained integrable \hs\ can be
related to the continuum formalism. The \K-\M\ transform of the
Virasoro-constrained KP hierarchy produces minimal matter, dressed with an
extra scalar, which we have further identified as arising from
BRST-invariance of highest-weight states of the topological algebra
\req{topalgebra}.
Although we have presented explicit derivations only for
level $l\!=\!2$,
a relation to the topological algebra according to a scheme
similar to the above
$\ket\Xi\rightarrow\ket\Upsilon$, can also be verified
explicitly for level
$l\!=\!3$ \cite{[GS3]}. While a complete proof valid for
an arbitrary level $l$ is lacking,
there is little doubt that it must exist.
Recall further that, as was shown for levels 2 and 3
in \cite{[S35],[GS]}, the \K-\M\ transform provides
a mapping (isomorphism?) between the appropriately dressed \de s
and \V\ \cs\ on the KP tau function\footnote{
The constraints appearing at level $l=3$ \cite{[GS]} were misinterpreted
originally as the W$^{(3)}$ ones; in fact, they do reduce to
the \V\ \cs\ \cite{[GS3]}.}.
 This leads us to conjecture that \V\ \cs\
on the KP \h\footnote{While our results apply directly to the KP \h, the
reductions of our scheme to the (constrained) generalized KdV hierarchies
\cite{[S29],[S35]}
would also be interesting to study.} are equivalent to the
BRST-invariance condition for highest-weight states of the topological
algebra.

We have shown that different constructions of the type $$\mbox{
matter + `Liouville' + ghosts}\Longrightarrow \mbox{topological\
algebra}\label{reduction}$$
are possible, depending on the prescription one
uses to dress the matter. While the more standard DDK case corresponds to
taking spin-2 ghosts, the spin-1 ghost system plays an important r\^ole as
well, since the corresponding
`breakdown' of the topological symmetry allows
one to write down the \de s in such a way that they lie in
the image of the \K-\M\ transform and in this way be equivalent to the \V\
 constraints on the KP \h.

Thus the topological algebra may be viewed as capturing the `invariant'
meaning of the theory, be it in the DDK formulation or in the guise of
constrained integrable \hs.
That is, splitting a specific ghost system out of
the topological
algebra depends on one's point of view: if one wishes to have
a worldsheet phase with {\it reparametrization\/} ghosts, {\sl then\/} the
standard DDK formalism
is recovered. On the other hand, following the $c=-2$
path, one ends up with matter dressed with a `Liouville' according to a
different prescription
\cite{[S35]} (cf. also \cite{[W-ground]}). The r\^ole
that the $c$-mode plays
in distinguishing between the two `reduced' pictures
suggests analogies with the existence of states which are different ghost
number `replicas' \cite{[W-ground],[KMS]} of representatives in BRST
cohomologies \cite{[LZ]}.
It is thus interesting to analyze the `invariant'
formulation with the topological symmetry `unbroken'.

As for the relation
\req{d(c)} between the matter central charge $d$ and the
topological central
charge $\ctop$, it is not so surprising that its values
exclude the region $1\!<\!d\!<\!25$, as that $\ctop$ is a {\it central
charge\/} as well.

Finally, let us note
that the direct relation of the `topological' approach
to the integrable formulation (\ie\ to constrained \hs), provided by the
\K-\M\ transform, can
be used together with the results of \cite{[VV-Apr90]}
to establish more direct links with the Landau--Ginzburg formulation. This
should have interesting implications for the topological Landau--Ginzburg
models \cite{[V]}.

\bigskip

\noindent
{\sc Acknowledgements.} We are grateful to S.~Pakulyak and V. Shadura for
 kind hospitality at the Institute of Theoretical Physics (Kiev). We thank
C.~Bachas,
E.~Corrigan, V.~Ya.~Fainberg, A.~Gorsky, S.~Kharchev, W.~Lerche,
A.~Lossev, J.~L.~Miramontes, A.~Mironov, A.~Morozov, P.~van~Nieuwenhuizen,
A.~Orlov, S.~Theisen I.~V.~Tyutin, M.~A.~Vasiliev and B.~L.~Voronov for
useful discussions.

\xpt


\begin{thebibliography}{55}
\parindent=0pt
\parskip=-5.5pt
\def\NPB{Nucl. Phys. B}
\def\PLB{Phys. Lett. B}
\def\MPLA{Mod. Phys. Lett. A}

\bibitem{[BK]} E.~Br\'ezin and V.~A.~Kazakov, \PLB 236 (1990) 144.
\bibitem{[DSh]} M.~R.~Douglas and S.~H.~Shenker, \NPB335 (1990) 635.
\bibitem{[GM]} D.~J.~Gross
and A.~A.~Migdal, Phys. Rev. Lett. 64 (1990) 127.
\bibitem{[D]} M.~R.~ Douglas, Phys. Lett. B238 (1990) 176.
\bibitem{[FKN1]} M.~Fukuma,
H.~Kawai and R.~Nakayama, Int. J. Mod. Phys. A6
(1991) 1385; {\sl Explicit Solution for $p$-$q$ Duality in Two-Dimensional
Quantum Gravity}, Univ. Tokyo preprint UT-582 (May 1991).
\bibitem{[DVV]} R.~Dijkgraaf, E.~Verlinde and H.~Verlinde, \NPB348 (1991)
435.
\bibitem{[KPZ]} V.~G.~Knizhnik,
A.~M.~Polyakov and A.~B.~Zamolodchikov, \MPLA
3 (1988) 819.
\bibitem{[Da]} F.~David, \MPLA3 (1988) 1651.
\bibitem{[DK]} J.~Distler and H.~Kawai, \NPB321 (1989) 509.
\bibitem{[S35]} A.~M.~Semikhatov, {\sl Solving Virasoro Constraints on
Integrable Hierarchies via the Kon\-tse\-vich-Mi\-wa Transform,\/}
hepth@xxx/9204063, \NPB, to appear.
\bibitem{[GS]} B.~Gato--Rivera and
A.~M.~Semikhatov,
\PLB 288 (1992) 38.
\bibitem{[K]} M.~Kontsevich, Funk. An. Prilozh. 25 (1991) N2, 50.
\bibitem{[W2]} E.~Witten, {\sl On the Konsevich Model and Other Models of
Two-Dimensional Gravity}, Princeton preprint IASSNS-HEP-91-24
 (July 1991).
\bibitem{[MS]} Yu.~Makeenko and G.~Semenoff, {\sl Properties of Hermitian
Matrix Model in External Field}, British Columbia Univ. prepr. 91-0329
(July 1991).
\bibitem{[KMMMZ]} S.~Kharchev, A.~Marshakov, A.~Mironov, A.~Morozov and
A.~Zabrodin,
{\sl Towards Unified Theory of Quantum Gravity,\/} Lebedev Inst.
preprint FIAN-TD-03-92 (Oct. 1991).
\bibitem{[GN]} D.~J.~Gross and M.~J.~Newman, {\sl Unitary and Hermitean
Matrices in an External Field II: The \K\ Model and Continuum \V\
Constraints}, Princeton preprint PUPT--1282 (1991).
\bibitem{[W-top]}E.~Witten, Commun. Math. Phys. 118 (1988) 411; \NPB 340
(1990) 281.
\bibitem{[EY]} T.~Eguchi and S.-K.~Yang, \MPLA4 (1990) 1653.
\bibitem{[BPZ]} A.~A.~Belavin, A.~M.~Polyakov and A.~B.~Zamolodchikov,
\NPB241 (1984) 333.
\bibitem{[FQS]}
D.~Friedan, Z.~Qiu and S.~Shenker, Phys. Rev. Lett. 52 (1984)
1575.
\bibitem{[DF]} Vl.~S.~Dotsenko and V.~A.~Fateev, \NPB 240 (1984) 312.
\bibitem{[W-Dec89]}E.~Witten, \NPB340 (1990) 281.
\bibitem{[DW-Feb90]} R.~Dijkgraaf and E.~Witten, \NPB342 (1991) 486.
\bibitem{[VV-Apr90]} E.~Verlinde and H.~Verlinde, \NPB348 (1991) 457.
\bibitem{[M]}E.~Martinec, \PLB217 (1989) 431.
\bibitem{[VW]}C.~Vafa and N.~P.~Warner, \PLB218 (1989) 51.
\bibitem{[W-ground]}E.~Witten, \NPB373 (1992) 187.
\bibitem{[KMS]}D.~Kutasov, E.~Martinec and N.~Seiberg, \PLB276 (1992)
437.
\bibitem{[LZ]} B.~Lian and G.~Zuckerman, \PLB254 (1991) 417; Commun. Math.
Phys. 135 (1991) 547.
\bibitem{[Mi]}T. Miwa, Proc. Japan Acad. Sci. 58 (1982) 9.
\bibitem{[Sa]} S.~Saito, Phys. Rev. D 36 (1987) 1819; Phys. Rev. Lett. 59
(1987) 1798.
\bibitem{[Di]} J.~Distler, \NPB324 (1990) 523.
\bibitem{[S10]} A.~M.~Semikhatov, Int. J. Mod. Phys. A4 (1989) 467.
\bibitem{[GO]} P.~G.~Grinevich and A.~Yu.~Orlov, {\sl Flag spaces in KP
Theory
and \V\ action on ${\mathrm det}\overline\d_j$ and Segal--Wilson Tau
Function\/}, Cornell Univ. prepr. CLNS 89/945.
\bibitem{[LVW]}W.~Lerche, C.~Vafa and N.~P.~Warner, \NPB324 (1989)
427.
\bibitem{[GS3]} B.~Gato--Rivera and A.~M.~Semikhatov, {BRST
Invariance, Constrained  Hierarchies, and the
Kontsevich--Miwa Transform}, in preparation.
\bibitem{[FMS]} D.~H.~Friedan, E.~J.~Martinec and S.~H.~Shenker, \NPB271
(1986) 93.
\bibitem{[S29]} A.~M.~Semikhatov, Nucl. Phys. B366 (1991) 347.
\bibitem{[V]} C.~Vafa, \MPLA6 (1991) 337.
\end{thebibliography}
\end{document}